\documentstyle[seceq,supplement,epsf,psfig]{ptptex}

\title{
Towards a Topological Mechanism of Quark Confinement 
}

\author{
Ernst-Michael {\sc Ilgenfritz}$^{1,2}$, Harald {\sc Markum}$^{3}$, 
Michael {\sc M\"uller--Preu\ss ker}$^{2}$, Wolfgang {\sc Sakuler}$^{3}$ 
and Stefan {\sc Thurner}$^{3}$
}

\inst{
$^{1}$Institute for Theoretical Physics, University of Kanazawa, Japan
\\
$^{2}$Institut f\"ur Physik, Humboldt--Universit\"at zu Berlin, Germany
\\
$^{3}$Institut f\"ur Kernphysik, Technische Universit\"at Wien, Austria
}

\recdate{
\today
}

\abst{
We report on new analyses of the topological and chiral
vacuum structure of four-dimensional QCD on the lattice. 
Correlation functions as well as visualization of monopole currents 
in the maximally Abelian gauge 
emphasize their topological origin and  
gauge invariant characterization. The (anti)selfdual character
of strong vacuum fluctuations is reveiled by smoothing.
In full QCD, (anti)instanton positions are also centers of the local 
chiral condensate and quark charge density. 
Most results turn out generically independent of the action 
and the cooling/smoothing method. 
}

\begin{document}

\maketitle

\section{Introduction}
As a matter of fact, nontrivial topology 
of non-Abelian gauge fields is a 
characteristic feature of non-perturbative QCD. An economical principle of 
nature leads us to
expect that its characteristic manifestation, instantons, may play  a
key role  in the still elusive mechanism of confinement, which is a very
peculiar property of QCD as well. Instantons appear in a semiclassical
path integral quantization of this theory while many people expect
that confinement itself is not a semiclassical phenomenon. If there is
a linkage, the part taken by the instantons can only be
an indirect one. What is the missing link? This question is almost as old
as the instanton solution. 
In a first attempt, 
after the instanton amplitude had been evaluated
by 't Hooft,\cite{thooft1}
Callan, Dashen and Gross \cite{cdg_bag}
tried to build this bridge in a more abstract way, considering instantons
contributing to the non-perturbative running of the QCD
coupling constant which, eventually, should give rise to a vacuum
that does not support quarks but could enclose color neutral bubbles 
(hadron bags).  
In this scenario instantons were 
topological objects by themselves, but their
relation to confinement would be of non-topological nature. The same
applies to attempts to obtain the string tension from Wilson loops
in an uncorrelated instanton gas.\cite{cdg_pot}  
Both approaches have failed to
produce confinement with realistic density and size distribution 
of instantons. Admittedly, the complicated interactions
of the semiclassical objects have been taken into account
only in a very crude way.
The density and size parameters are more or less known, from the
resolution of the axial $U(1)$ problem via   
the Witten-Veneziano formula,  
from low energy hadronic spectroscopy (chiral symmetry breaking) and, 
with instanton sizes differing within a factor of two, 
also from lattice investigations addressing directly the
topology of gauge field configurations. 

For a long time, the conclusion was  
that instantons are unrelated to confinement. 
An understanding of confinement seemed decoupled from hadron physics and not 
really needed to 
understand the latter being 
described exclusively by physics at the instanton scale. 
Indeed, the topological charge $Q=\pm 1$ of isolated instantons  
gives rise to fermionic zero-modes 
via the Atiyah-Singer index theorem.
In a dense liquid of instantons,
they are split into a band of quasi-zero-modes. The detailed interaction
of instantons, their density and overlap, determines the chiral
symmetry breaking and the long-range propagation in light hadronic 
channels.\cite{shuryak}  From this perspective, 
confinement appears to be a somewhat esoteric issue. 

On the other hand, there is a viable explanation of confinement by an
effective
dual Abelian Higgs theory,\cite{suzuki}  
with the charged scalar field being  
in the condensed phase, where confinement is realized
by an area law of t'Hooft loops instead of Wilson loops. If this scalar
field represents, by particle-field duality, magnetic charges
of the theory we are studying, then it is tempting to look for 
signals of their condensation in usual lattice QCD configurations.
Various gauges have been tested 
\cite{polikarpov}
to extract {\it the relevant} degrees of freedom
at long distances, being Abelian and giving rise
to magnetic charges in an appropriate projection, 
and to provide evidence for condensation. 
While the effective
action of {\it this reduced} Abelian gauge field theory 
coupled to charges and monopoles 
is probably too complicated to be {\it effective}, the reduced action
of monopoles alone can be put into relation 
with the above mentioned {\it dual} Abelian Higgs model.\cite{kato}  
The phenomenological effective dual Ginzburg-Landau 
theory can also account for 
chiral symmetry breaking
and restoration.\cite{toki} 
It seems, that the concept
of the dual superconductor mechanism \cite{dualsuper} can be worked out 
in this framework, and the actual agents of confinement are monopoles
with respect to the Abelian subgroup $U(1)^{N-1}$ of
color $SU(N)$.

If, in this context, instantons support confinement, they must 
provide the basic mechanism which helps monopoles to condense.
A few years ago, it has been demonstrated that Abelian monopole currents
(identified in the maximally Abelian gauge)
and topological density are highly correlated 
quantities, in Monte Carlo ensembles \cite{wien} as well as for 
instantons and instanton-antiinstanton 
pairs.\cite{brower}  
It has been demonstrated that the Abelian monopoles carry also electric
charge. \cite{borny} 
This observation lends support to dyonic vacuum models.\cite{simonov} 
All this has been taken as indication that Abelian monopoles as agents of the
confinement mechanism might have a topological
origin such that finally both approaches can be united.
Reconstructing gauge fields from instanton ensembles identified on
genuine lattice gauge field configurations fails to reconstruct
the string tension.\cite{degrand} 
Some important parameters characterizing instanton
liquids as far as confinement is concerned
seem to be unidentified so far. 
It has been conjectured, that not only instantons but monopoles, too, are 
related to chiral symmetry breaking.\cite{miyamura95} In contrast to
the case of confinement, here is even 
evidence on the lattice that the reconstruction of gauge fields from the
monopole degrees of freedom successfully describes the behavior of the
near-to-zero-modes of the fermions.\cite{japaner} Chiral symmetry breaking
is a collective effect. It seems that the missing
information about the large scale structure of the instanton liquid is 
encoded in the monopole degrees of freedom.

Today, in distinction to the times of the 
pioneering work by 
Callan, Dashen and Gross,   
there exists a multitude of tools within 
the lattice approach to address afresh
the question of confinement due to instantons and to come, 
hopefully, to a definite
answer. After all, besides all approximate   
non-perturbative working schemes
for (non-supersymmetric) QCD, lattice simulation is the
only method based solely on first principles.
In this contribution we give further 
support for the ideas above.  We discuss new tools for the investigation of the
instanton-confinement issue.
We demonstrate the way how instantons and monopoles 
coexist 
on individual 
gauge field configurations, by extracting correlation functions 
from individual fields and by direct visualization, both in the 
confinement and in the 
deconfinement phase. 
We present selected results obtained by a 
Wilson renormalization group 
motivated smoothing method \cite{hasenfratz,berlin96,berlin98} 
and point out the dominance of locally (anti)self dual gauge fields.
We report also of  
new results on local 
correlations between the quark condensate and the quark charge density 
on one hand and the topological charge 
and the local monopole current density on the other, illustrating the
mechanism of chiral symmetry breaking. \\

\section{Observables and Techniques}

There exist several definitions of the Pontryagin number and its
density on a 
Euclidean lattice.
One class are 
the so-called field theoretic definitions which
discretize the topological charge density in the continuum,
$
q(x)=\frac{g^{2}}{32\pi^{2}} \epsilon^{\mu\nu\rho\sigma}
\ \mbox{\rm Tr} \ \left( F_{\mu\nu}(x) F_{\rho\sigma}(x) \right) \ ,
$
in the following two ways:\cite{divecchia}
\begin{equation}
\label{eq:qdefinition}
 q^{(P,H)}(x)=-\frac{1}{2^{4}32\pi^{2}}
\sum_{\mu,\ldots =\pm 1}^{\pm 4}
\tilde{\epsilon}_{\mu\nu\rho\sigma} \mbox{\rm Tr} \ O_{\mu\nu\rho\sigma}^{(P,H)}
,
\end{equation}
with
\begin{equation}
\label{eq:plaquette_oriented}
O_{\mu\nu\rho\sigma}^{(P)} = U_{\mu\nu}(x) U_{\rho\sigma}(x) \ 
\end{equation}
the plaquette prescription, and by the hypercube prescription
\begin{eqnarray}
\label{eq:hypercube_oriented}
O_{\mu\nu\rho\sigma}^{(H)} &=&
            U_{\mu}(x) 
	    U_{\nu}(x\!+\!\hat\mu) 
            U_{\rho}(x\!+\!\hat\mu\!+\!\hat\nu)
            U_{\sigma}(x\!+\!\hat\mu\!+\!\hat\nu\!+\!\hat\rho) \nonumber \\
 & \times & U^{\dagger}_{\mu}(x\!+\!\hat\nu\!+\!\hat\rho\!+\!\hat\sigma)
            U^{\dagger}_{\nu}(x\!+\!\hat\rho\!+\!\hat\sigma)
            U^{\dagger}_{\rho}(x\!+\!\hat\sigma) 
	    U^{\dagger}_{\sigma}(x).
\end{eqnarray}
Besides of this, there 
is a geometric definition \cite{luescher} where 
integrals over transition functions between local gauge patches are 
defining local contributions to the topological charge.

To investigate monopole currents we project $SU(N)$
onto its Cartan subgroup extracting Abelian degrees of freedom in a way
that leaves the Abelian $U(1)^{N-1}$ gauge symmetry intact.
We employ the 
so-called maximally Abelian gauge.\cite{kronfeld} This gauge is
the most likely one to guarantee that the Abelian projection captures
long range physics encoded in the gauge field configurations. 
Monopole charge currents can then be defined on the
dual lattice which
are both quantized and topologically conserved.
For purposes of measuring correlations, the monopole currents 
$m(x,\mu)$ 
are locally summed into a rotational invariant  
monopole density 
\begin{equation}
\label{eq:magn_current_dens}
\rho_m(x) = \frac{1}{4} \sum_{\mu} | m(x,\mu) | .  
\end{equation}

A local quark condensate can be defined by
inverting the fermionic matrix $M$ of the QCD action and taking the
trace over all but the space-time indices of $M^{-1}$, 
\begin{equation}
\label{eq:psibarpsi}
\bar \psi \psi (x) \equiv  \mbox{tr} M^{-1} . 
\end{equation}
Another observable of interest is the local fermionic charge 
density 
\begin{equation}
\label{eq:psidaggerpsi}
\psi^{\dagger} \psi(x) \equiv  \mbox{tr} \gamma_4 M^{-1} .
\end{equation}
Since topological objects with opposite sign are equally probable in the
quantum ensemble,
we measure the correlation of the monopole density and the local quark
condensate 
with the absolute values 
of the topological charge density and of the fermionic charge density.

\begin{figure}[t]
\begin{tabular}{cc}
\hspace{0.3cm}  Confinement & \hspace{0.3cm}  Deconfinement\\
\epsfxsize=6.7cm\epsffile{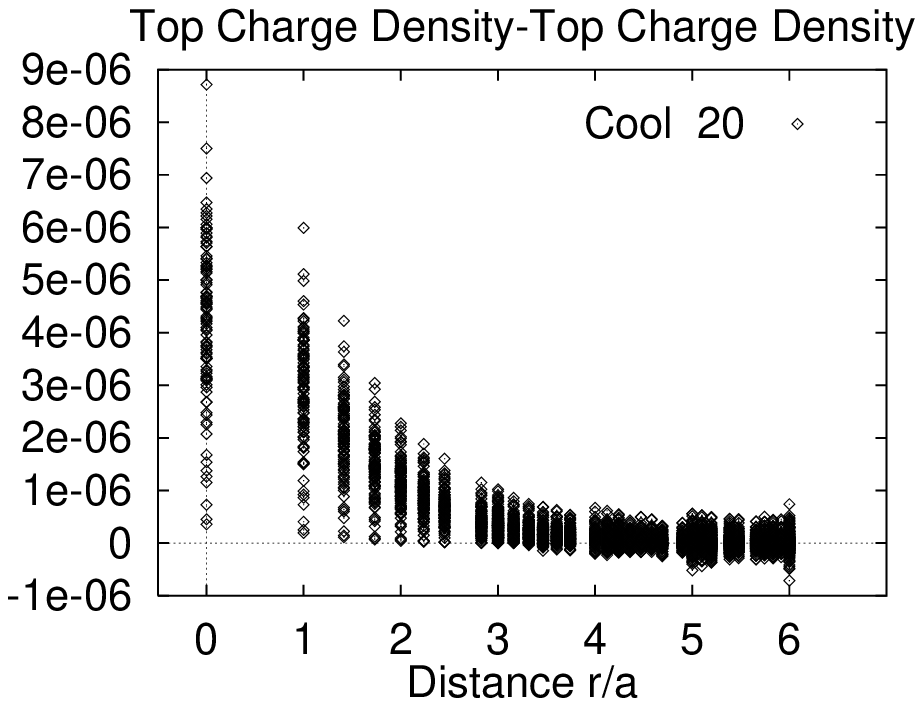}& 
\epsfxsize=6.7cm\epsffile{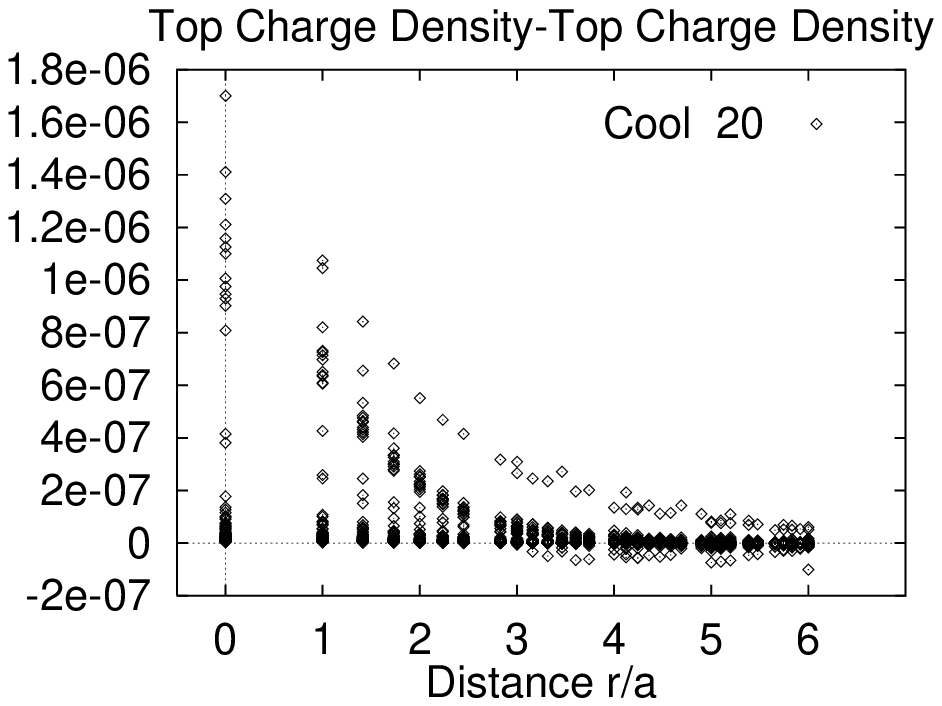}\\
\end{tabular}
\vspace{-0.1cm}
\caption{
Correlation functions of the topological charge density 
in 100 individual  configurations representing both phases 
of pure $SU(2)$ gauge theory. Measured after 20 cooling steps. } 
\end{figure}
%
\begin{figure}[t]
\begin{tabular}{cc}
\hspace{0.3cm}  Confinement & \hspace{0.3cm}  Deconfinement\\
\epsfxsize=6.7cm\epsffile{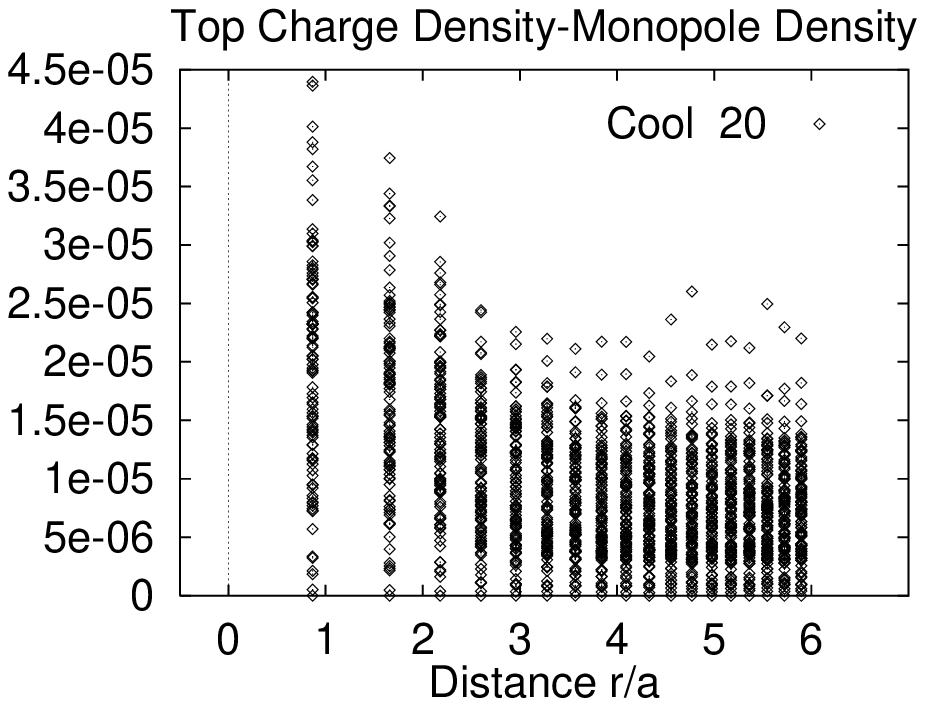}& 
\epsfxsize=6.7cm\epsffile{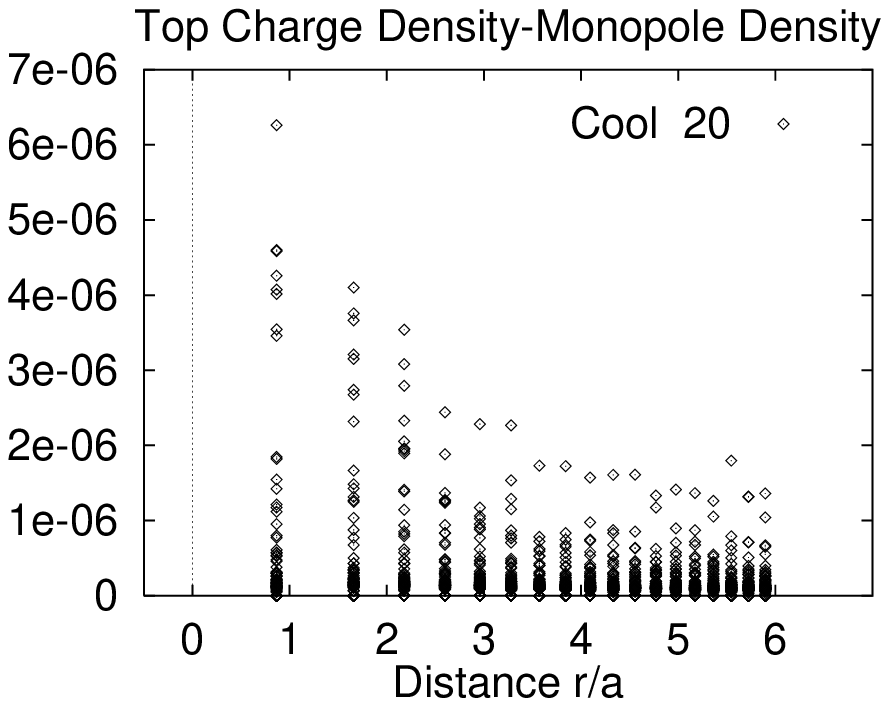}\\
\end{tabular}
\vspace{-0.1cm}
\caption{
The $\rho_m|q|$-correlation functions of individual configurations 
are displayed for both phases. All configurations 
with nonvanishing 
$qq$-correlation 
give rise to a nontrivial $\rho_m|q|$-correlator.} 
\end{figure}
%
Some of the correlators require the suppression of ultraviolet
quantum fluctuations. 
In most of our studies we have employed the controlled (small step-size)
cooling technique which, when applied to
the $SU(3)$ case, uses the 
Cabibbo-Marinari method (cooling within the $SU(2)$
subgroups). 
In the confinement phase, monitoring the string tension, we 
keep control to what extent
the cooled ensemble still reproduces the vacuum properties. 
In the deconfinement phase, the same conditions of mild cooling are applied. 
Finally, however, the cooling method drives all
gauge field configurations into classical configurations. 
In a theoretically uncontrolled way important 
physical information like instanton 
sizes, relative positions and color orientation will be finally erased.
Therefore, more recently, we have supplemented our methods by a 
renormalization group
motivated smoothing technique.\cite{berlin96,berlin98} Being a specific 
inversion of the block spin transformation from a fine to a
coarse lattice, this
method aims to reconstruct the smoothest interpolation for any given lattice
configuration. This is achieved by a constrained minimization of
a classically perfect action. Since the coupling parameters of this
action are determined with the help of inverse blocking, this
method is theoretically self-consistent. The smoothed configurations
conserve the physics at scales above the resolution of the coarser lattice. 

\section{Instantons and Monopoles} 
 
Our search for the interrelation between topological charge
density and magnetic charge currents in the quantized vacuum starts with the 
measurement of the correlations
between the corresponding operators defined above. 
Going beyond previous correlation measurements, we have now
evaluated the correlation functions for individual configurations,
for instance
\begin{equation}
\label{eq:correlator}
 C_{qq}(r)=\frac{1}{L_s^3~L_{\tau}}\sum_{x,y} q(x) q(y) \delta(r-x+y) .
\end{equation}
We have analyzed lattice gauge field configurations created 
with the pure $SU(2)$ standard Wilson action  
on a $12^{3} \times 4$ lattice both in the confinement
and deconfinement phase at 
$\beta=2.25$ and $2.4$, respectively.

Fig.~1 presents the 
correlation function of the topological charge $q(x)$ 
with itself
and Fig.~2  the $\rho_m |q|$-correlation function
for 100 different configurations. The measurements refer to cooled
configurations after 20 cooling steps where in the confinement 
the string tension of the ensemble still amounts to about 
$80$\%  of the uncooled string tension.
In the confinement phase, the configurations differ strongly in
the amplitudes (not in the shapes) of the 
$qq$-correlation functions.
This reflects the big width of the multiplicity distribution in the number of
instantons and antiinstantons.
Also the corresponding correlators between monopole and topological
density show many 
different amplitudes. 
They decrease  towards 
the nonzero cluster value which is not subtracted and  represent the
nontrivial charge content of the particular configuration. 
In the deconfinement phase only approximately  
$15$\% of the configurations have nontrivial 
$qq$-correlation 
functions (an amplitude larger  than $10$\% of the biggest in the sample). 
All these configurations give  rise to 
a nontrivial $\rho_m |q|$-correlation. This illustrates how the correlation 
between monopoles and topological charge measured previously by averaging over 
the (cooled or uncooled) quantum ensemble
manifests itself for individual configurations.
It shows that the amplitude is correlated with the topological activity 
$A=\sum_x |q(x)|$ which becomes  
an approximation for the number of instantons and antiinstantons 
if the (cooled) configurations are sufficiently smooth. 
Concerning the shapes of both correlation functions 
an additional analysis could clarify
whether their approach to the cluster value is exponential
with a screening mass eventually depending  
on the topological content of the configuration.  

As an instructive example, we 
visualize in Fig.~3 the topological charge distribution
together with the monopole currents
for a configuration with net topological charge $Q=0$
taken from the 
ensemble describing the deconfinement phase.
We show clusters of topological charge and the
accompanying monopole loops, in two time-slices in the upper row, and in two
fixed space-slices in the lower. 
For any value of the topological charge density $q(x) > 0.005$ a light dot
and for $q(x) < -0.005$ a dark dot is plotted.
Monopole loops are represented by thick lines.
This configuration contains only one instanton and one antiinstanton. 
In the deconfinement phase, monopole loops 
are purely time-like, almost static
and must exist in monopole-antimonopole pairs. Here, each (anti)monopole
loop tends to pass through an (anti)instanton which extends throughout all
four time-slices.
In the confinement phase, such pictures look qualitatively similar except that 
also space-like monopole currents exist and monopole loops are closed
in all Euclidean directions.\cite{wien}

\begin{figure}[!th]
\begin{center}
\begin{tabular}{cc}
\ & \\
\large t = 2 fixed & \large t = 4 fixed  \\
\ & \\
\epsfxsize=6cm\epsffile{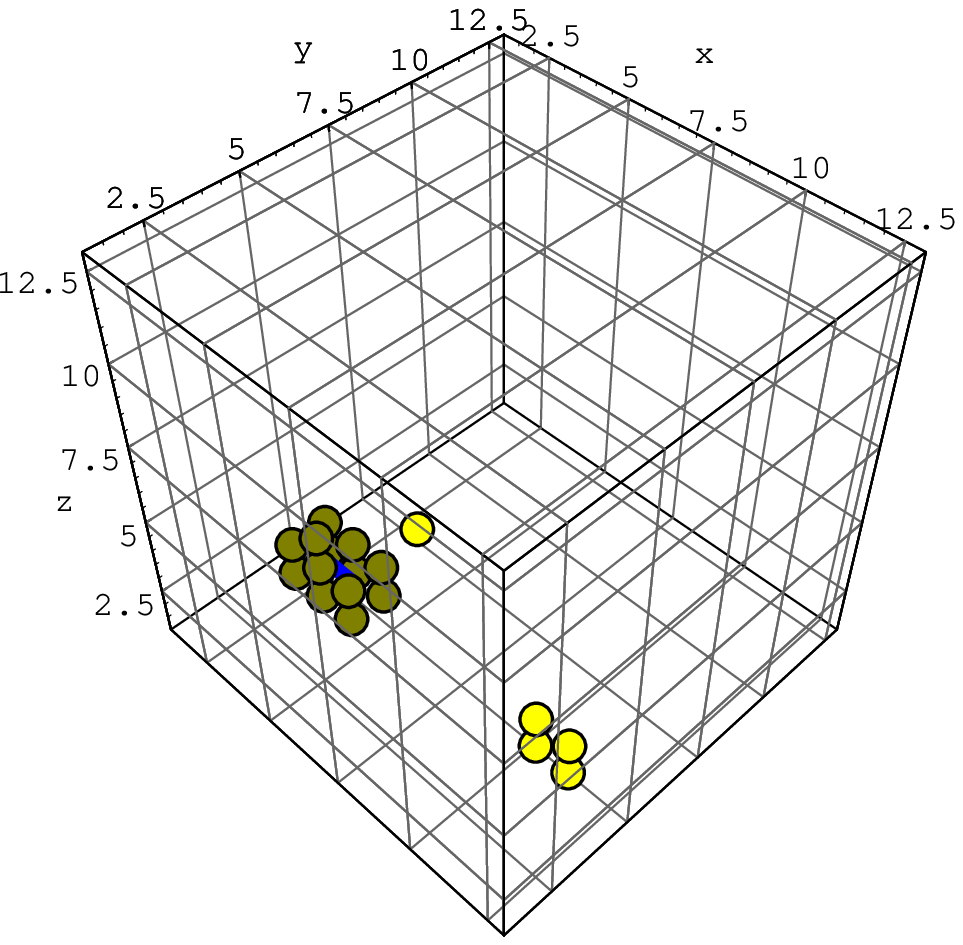} \vspace{0.0cm} & 
\epsfxsize=6cm\epsffile{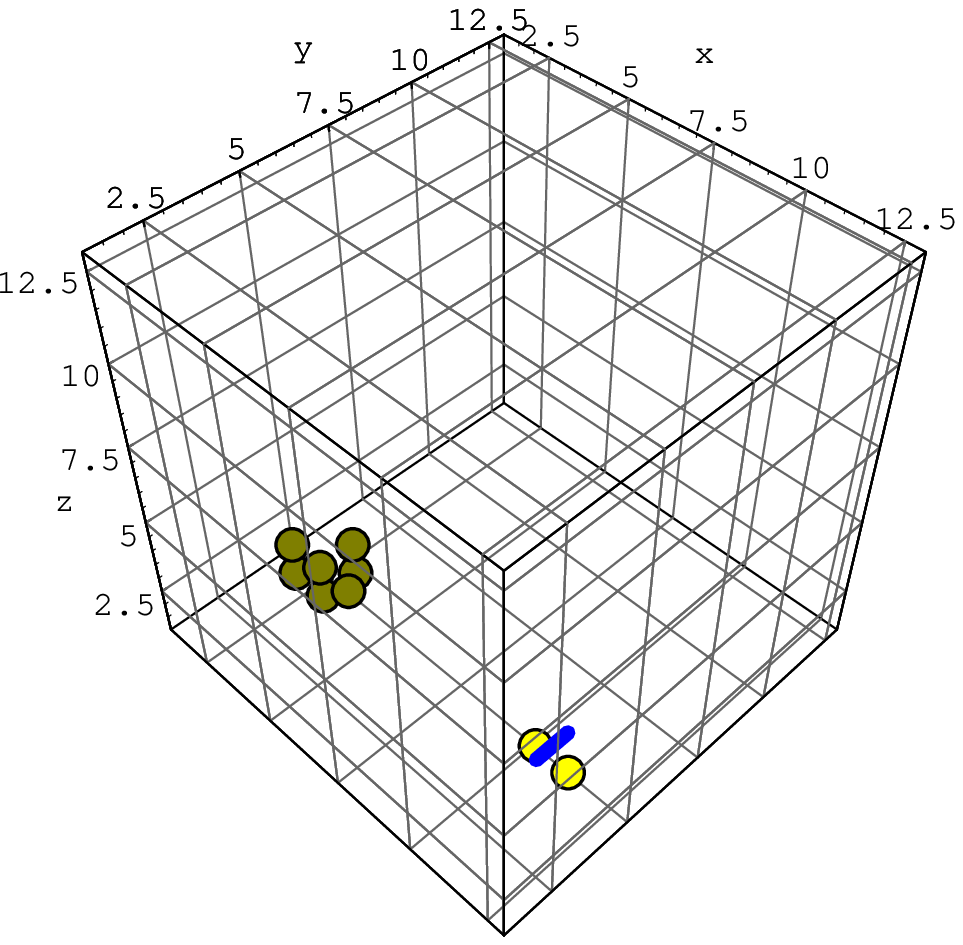} \vspace{0.0cm} \\
\ & \\
\large x = 4 fixed & \large x = 11 fixed  \\
\ & \\
\epsfxsize=6cm\epsffile{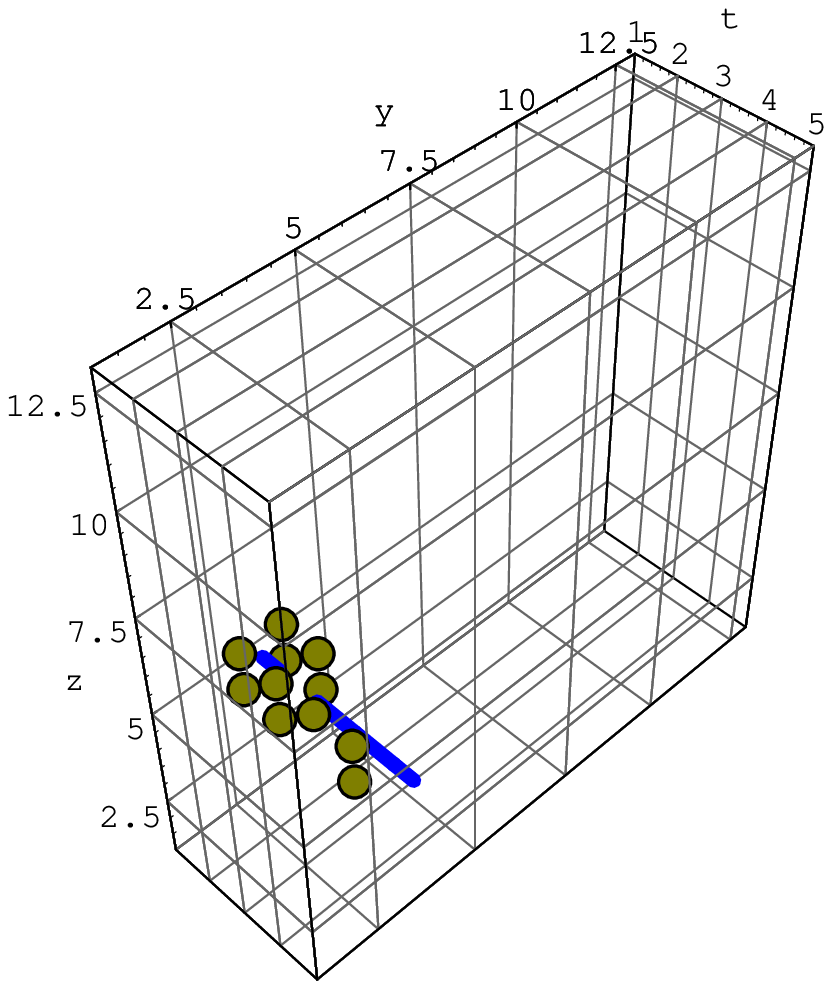} \vspace{0.0cm} & 
\epsfxsize=6cm\epsffile{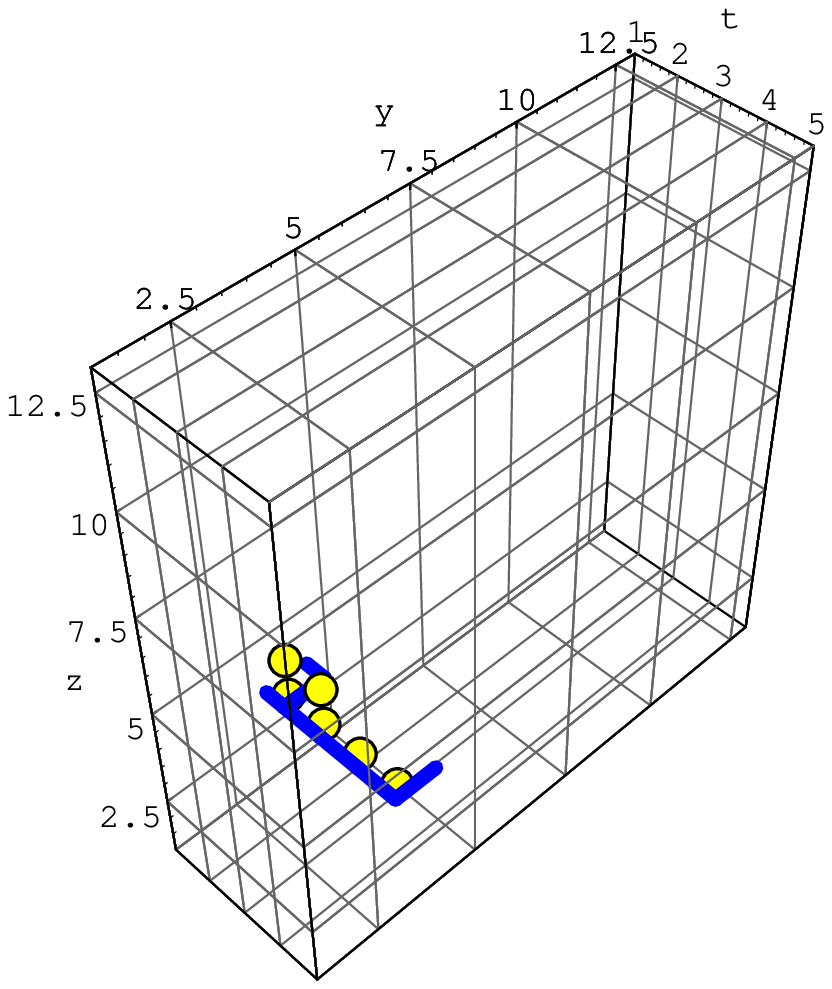} \vspace{0.0cm} \\
\end{tabular}
\end{center}
\caption{
Two time-slices (upper row) and two fixed space-slices (lower row)
visualizing a four-dimensional
configuration (after 20 cooling steps) taken from the deconfinement 
sample, 
containing an instanton-antiinstanton and a monopole-antimonopole
pair.
The dots show where the topological charge density $|q(x)| > 0.005$.
Thick lines represent monopole world lines.
}
\end{figure}

\begin{figure}[t]
\begin{center}
\begin{tabular}{cc}
\epsfxsize=6.6cm\epsffile{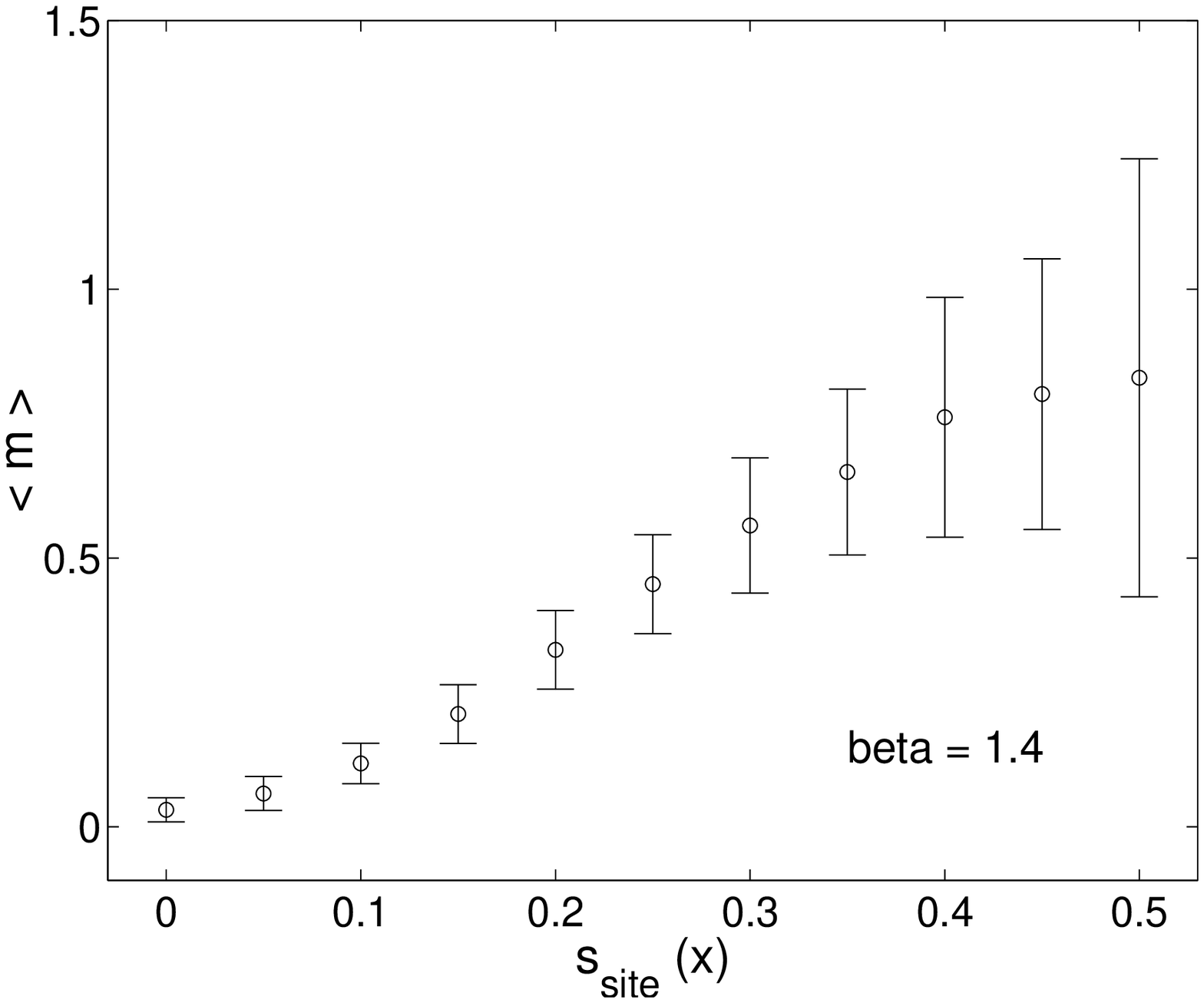}  &
\epsfxsize=6.3cm\epsffile{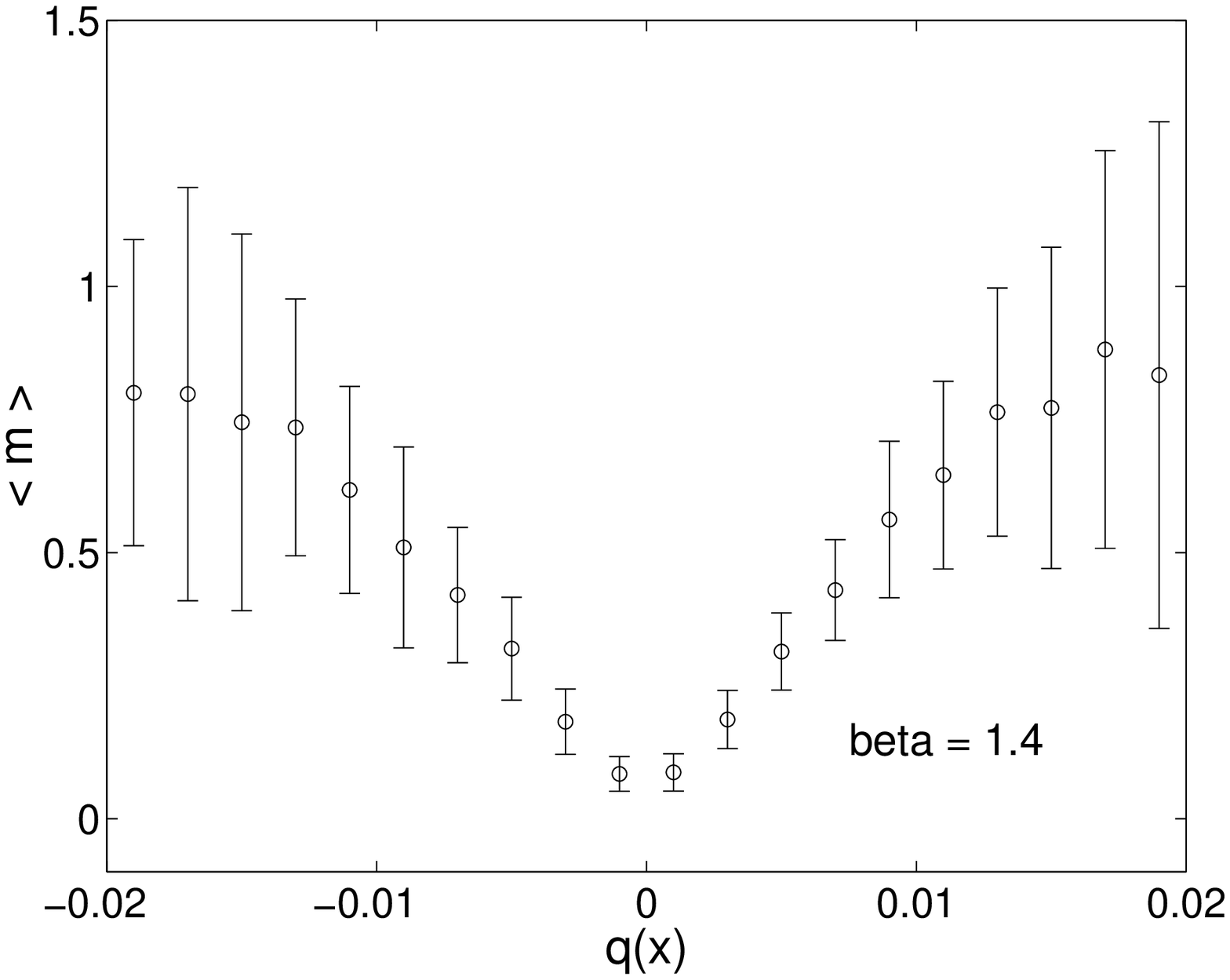}  
\end{tabular}
\end{center}
\caption{
Average occupation number of magnetic monopoles $\langle m \rangle$
on the dual links nearest to a lattice point $x$,
in dependence on the local action  $s_{site}(x)$ (left plot)
or the topological density $q(x)$ (right plot) for minimally smoothed
configuration.
The figure depicts the situation in the confinement phase at $\beta=1.4$,
being similar in the deconfinement phase.
} 
\end{figure}
These observations are corroborated by our recent studies using an
$SU(2)$ fixed-point action and
a renormalization group based  
smoothing method \cite{berlin96,berlin98}
instead of cooling.
This method allows to study the structure of the Yang-Mills vacuum
without cooling artefacts. The truncated fixed-point action is
a combination of plaquette and tilted 6-link loops and contains 
the traces up to the fourth power. 
Based on the smoothing method
we are able to give more detailed evidence for the local correlation 
of gauge invariant
quantities (topological charge density and action density)
with the presence of monopole currents. 

Fig.~4 shows for  smoothed configurations 
the average occupation number of magnetic monopoles $\langle m \rangle$
on the $32$ nearest dual links
as a function of  the local action $s_{site}(x)$ concentrated around a given lattice point.
A sample consisting of $50$ configurations at
$\beta=1.4$ in the confinement phase 
has been analyzed for this plot.
Apart from the change
in the range of the distribution of the
local action values with $\beta$,
the same dependence is observed 
in the deconfinement phase.
Fig.~4 additionally shows the dependence of the average
number of monopole currents on the local topological charge density
for the same sample of configurations.
There are practically no lattice points with
topological density $|q(x)| > 0.05$, irrespectively of $\beta$.
A physically important point 
is the increase of the average local density of monopole currents
from  small action density $s_{site}(x)$ or topological density $|q(x)|$
and the saturation 
for larger values of $s_{site}(x)$ and $|q(x)|$, respectively.
The increase of the variance with $s_{site}$ or $|q|$ reflects the
smaller number of lattice points with large action
or topological density.
These results have been obtained using the maximally Abelian gauge
for the Abelian projection and the subsequent 
localization of the Abelian monopoles. 
A preliminary study using the so-called Polyakov gauge (diagonalizing the
Polyakov loop for all space-time lattice points) 
indicates that the corresponding Abelian 
monopoles carry  less action. 

\begin{figure}[!thb]
\label{fig:joint_s_and_q}
\begin{tabular}{c}
\epsfxsize=11.0cm\epsffile{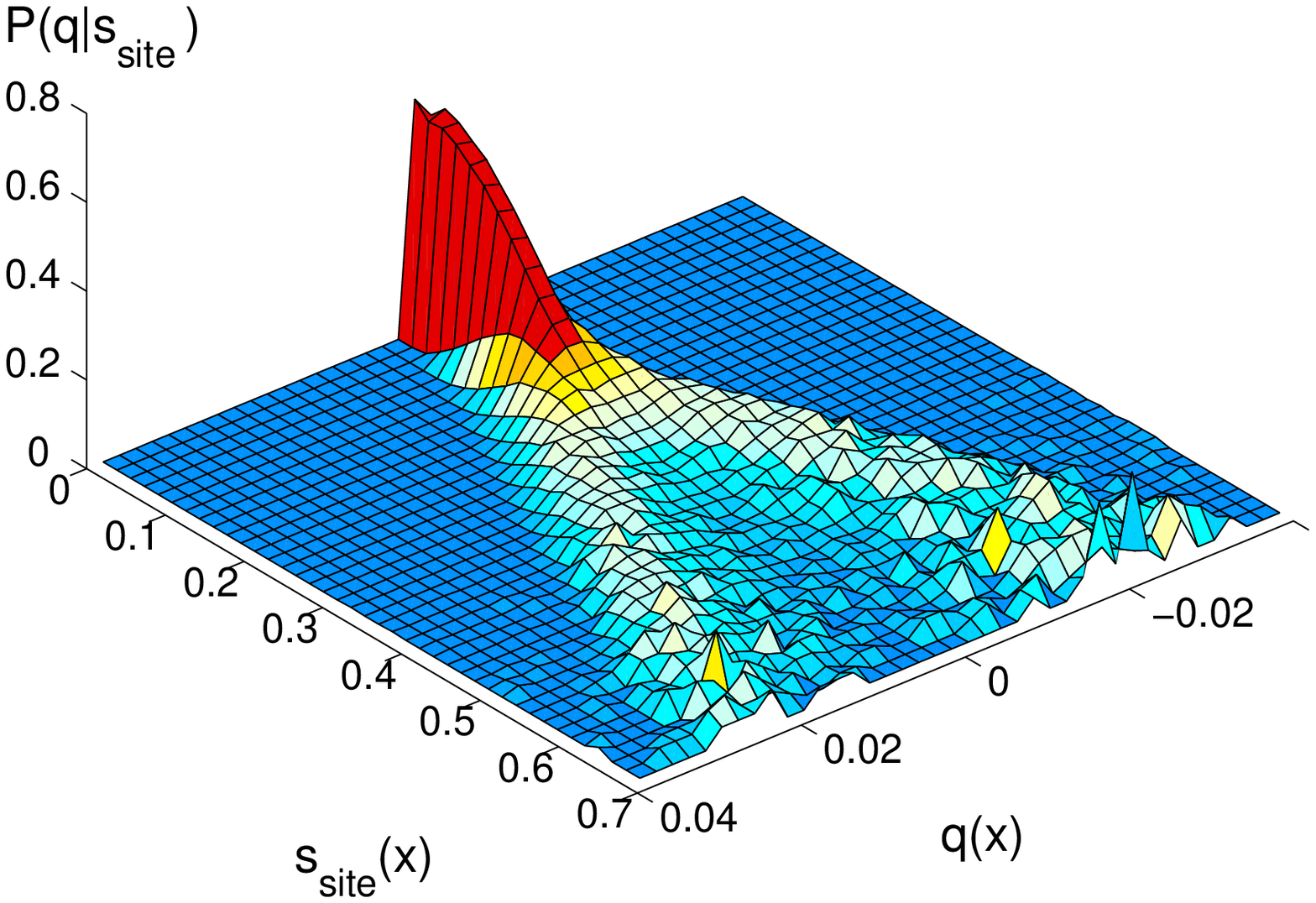}\\
\end{tabular}
\caption{  
Probability distributions for finding a local topological
charge $q(x)$ at a lattice site $x$
where the local action is found to be equal to $s_{site}(x)$.
$50$ independent confinement configurations obtained
for a fixed-point action at $\beta=1.4$
have been analyzed and the
$q$ distributions are separately normalized for each value of $s_{site}$. 
 }
\end{figure}
$3$-dimensional visualizations like Fig.~3 as well as $2$-dimensional
contour plots in our recent analysis based on the 
smoothing method \cite{berlin98}
give clear evidence for clustering of the topological density
at a scale of several lattice spacings
within any possible section of the lattice. On the minimally 
smoothed configurations, reproducing the quantum configurations  
at any scale above two  lattice spacings, the topological density
is a continuous function of all Euclidean coordinates.
However, a parametrization as classical instantons
is hardly possible in the confinement phase if no more sophisticated
smoothing techniques \cite{degrand} are applied which are 
biased in some way or another. 
Taking the action instead of the topological density, the same difficulty 
is encountered. 
Nevertheless, one can establish the fact that
strong gauge fields are locally nearly selfdual or antiselfdual. 
A local analysis shows that approximately selfdual or antiselfdual
domains are clearly preferred in both phases
when  the local action exceeds a threshold value of
$s_{site}(x) \approx 0.3$. This is demonstrated by 
the probability distribution of the local topological charge at
given local action density in
Fig.~5. In other words, smoothing reveals that color-electric
and -magnetic fields, if they are strong enough, locally 
tend to be mutually aligned in color and $3d$ space.
Therefore we must conclude 
that the (anti)instantons forming a dense fluid become strongly deformed
due to interactions and
under the influence of quantum fluctuations.
We find for all $\beta$-values that
the (plaquette oriented) topological density is locally bounded 
by the action density
\begin{equation}
2 \pi^2 ~|q(x)| \leq  s_{site}(x) \, .
\end{equation}
This is not unexpected and only confirms that the naive 
lattice expression (\ref{eq:plaquette_oriented}) 
for the topological charge density is a
good operator for smoothed configurations. This is supported also
by the correlation between the naive total charge and the geometric
charge \cite{berlin96} giving rise to 
a charge renormalization factor near to one.

\section{Local Chiral Condensate and Quark Charge Density}

Here we present results on the correlations of some fermionic
observables with  the topological density and the monopole density. 
The simulations have been performed for $SU(3)$ full QCD with dynamical
Kogut-Susskind fermions using the pseudofermionic method 
on an $8^{3} \times 4$ lattice.  
The configurations have been generated in the confinement phase 
at $\beta=5.2$ and  with 
$3$ flavors of degenerate mass $m=0.1$. 
When the configurations have been cooled
all observables including the fermionic ones 
have been reevaluated for the ensemble of cooled gauge fields. 
It must be stressed that
cooling here, as in the previous investigations for quenched QCD, 
denotes a controlled suppression
of ultraviolet gluonic quantum fluctuations with respect to the
Wilson action, {\it without} taking the
fermionic contribution to the 
effective gluonic action into account.
\begin{figure}[!b]
\label{corr}
\begin{tabular}{ccc}
\hspace{-9mm} \epsfxsize=5.2cm\epsffile{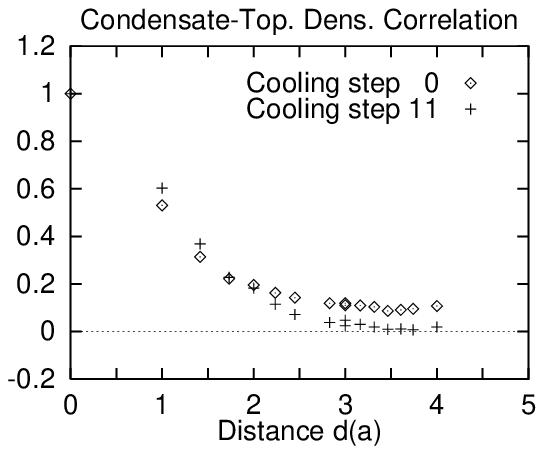}& 
\hspace{-9mm} \epsfxsize=5.2cm\epsffile{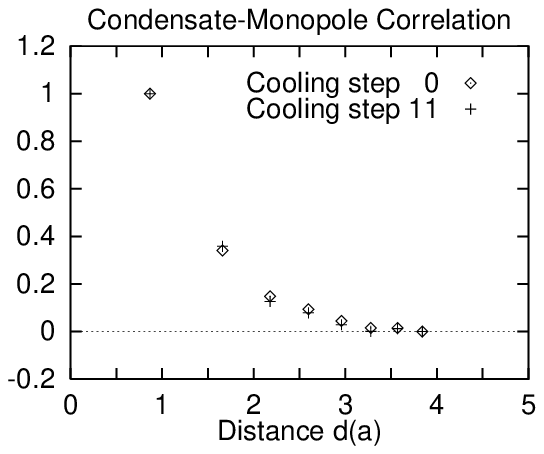}& 
\hspace{-9mm} \epsfxsize=5.2cm\epsffile{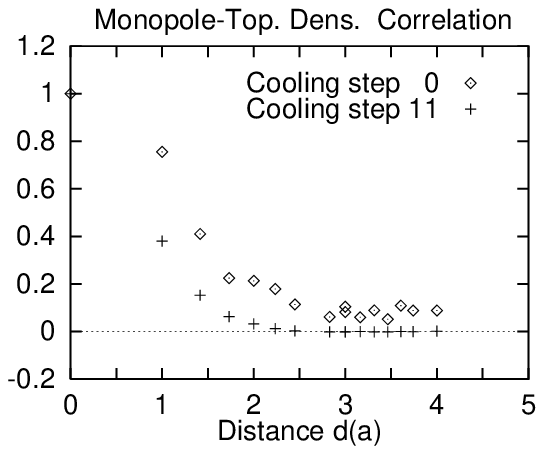}\\
\end{tabular}
\caption{Correlation functions of the local quark condensate
with the topological density squared  
and with the monopole density
(left two plots) and between topological and monopole density. 
The average is from  a non-quenched simulation 
in the confinement phase. 
Error bars  are comparable to the size of the symbols.
}
\end{figure}

Fig.~6 shows results for connected correlation functions of  
$\bar \psi \psi(x)$ with $\rho_m(x)$ and $ q^2(x)$, 
averaged over $1000$ configurations and normalized at the 
smallest possible distance. 
All correlation functions are clearly non-vanishing over 
distances bigger than two lattice spacings, even for the uncooled ensemble.
The correlation of the local quark condensate and the topological charge 
density is  not unexpected,\cite{hands}
and it does not vanish for the largest separations.
The long-range correlation becomes reduced with cooling, because   
fermionic zero-modes of individual instantons start dominating 
locally in the Kogut-Susskind fermion propagator $~M^{-1}$.\cite{ilmpss} 
The correlation between the condensate
and the monopole density turns out, however, rather cooling independent 
whereas the 
topological-charge monopole correlation 
decreases slightly with cooling. 

\begin{figure}
\begin{tabular}{cc}
0  Cooling steps & 5  Cooling steps\\
\psfig{figure=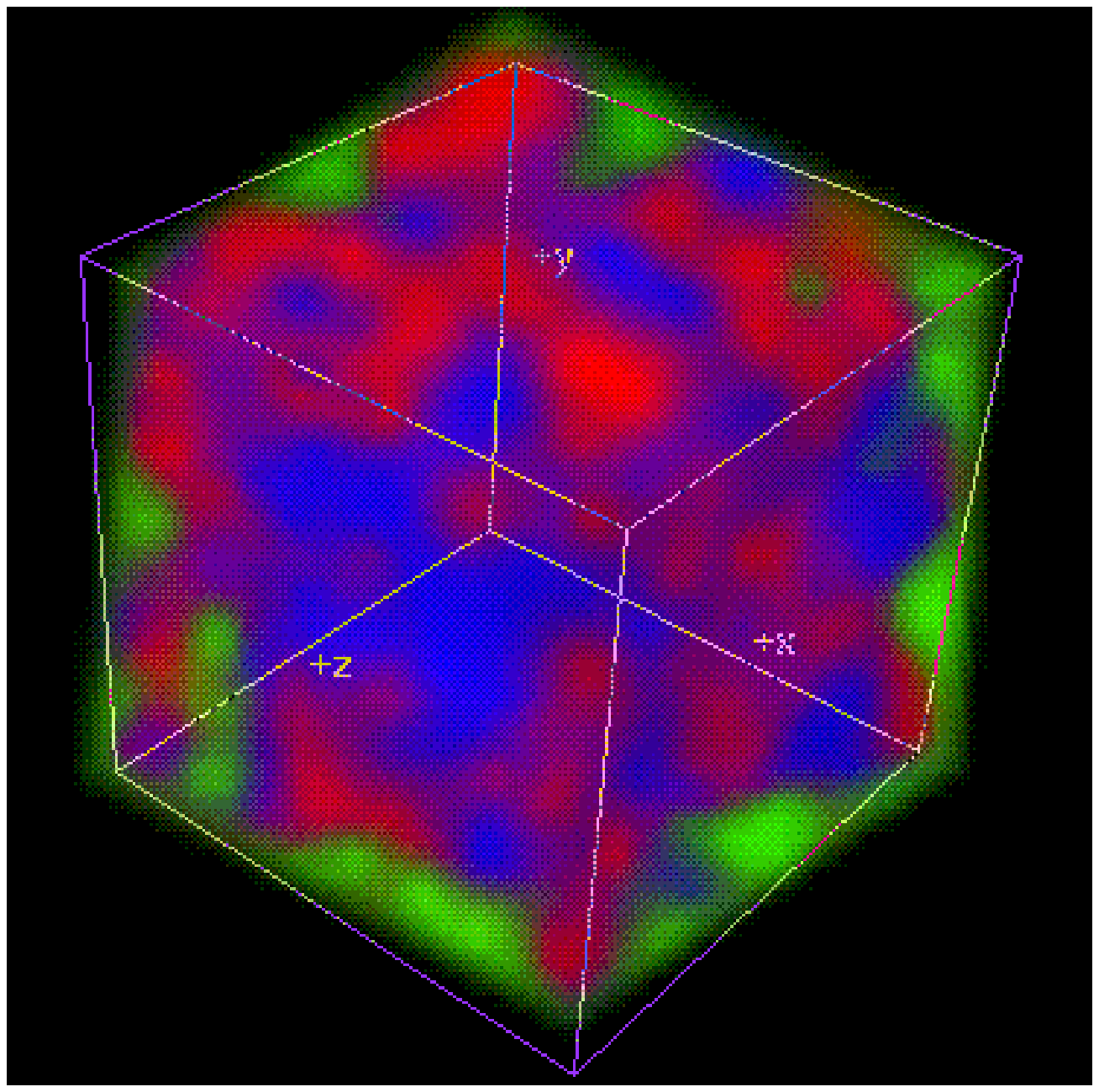,height=6.3cm,width=6.3cm} &
\psfig{figure=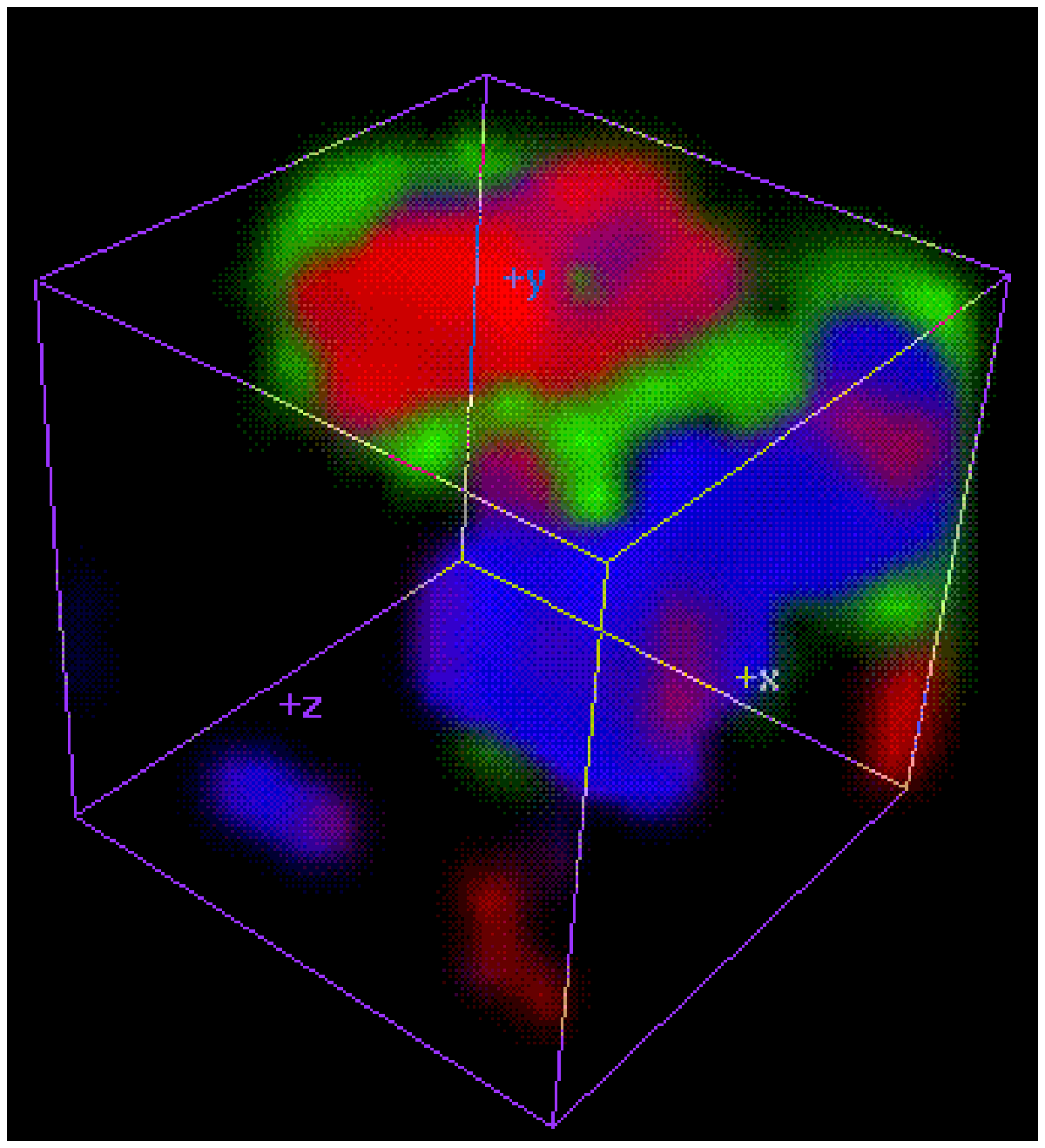,height=6.3cm,width=6.3cm} \\
10 Cooling steps & 15 Cooling steps\\
\psfig{figure=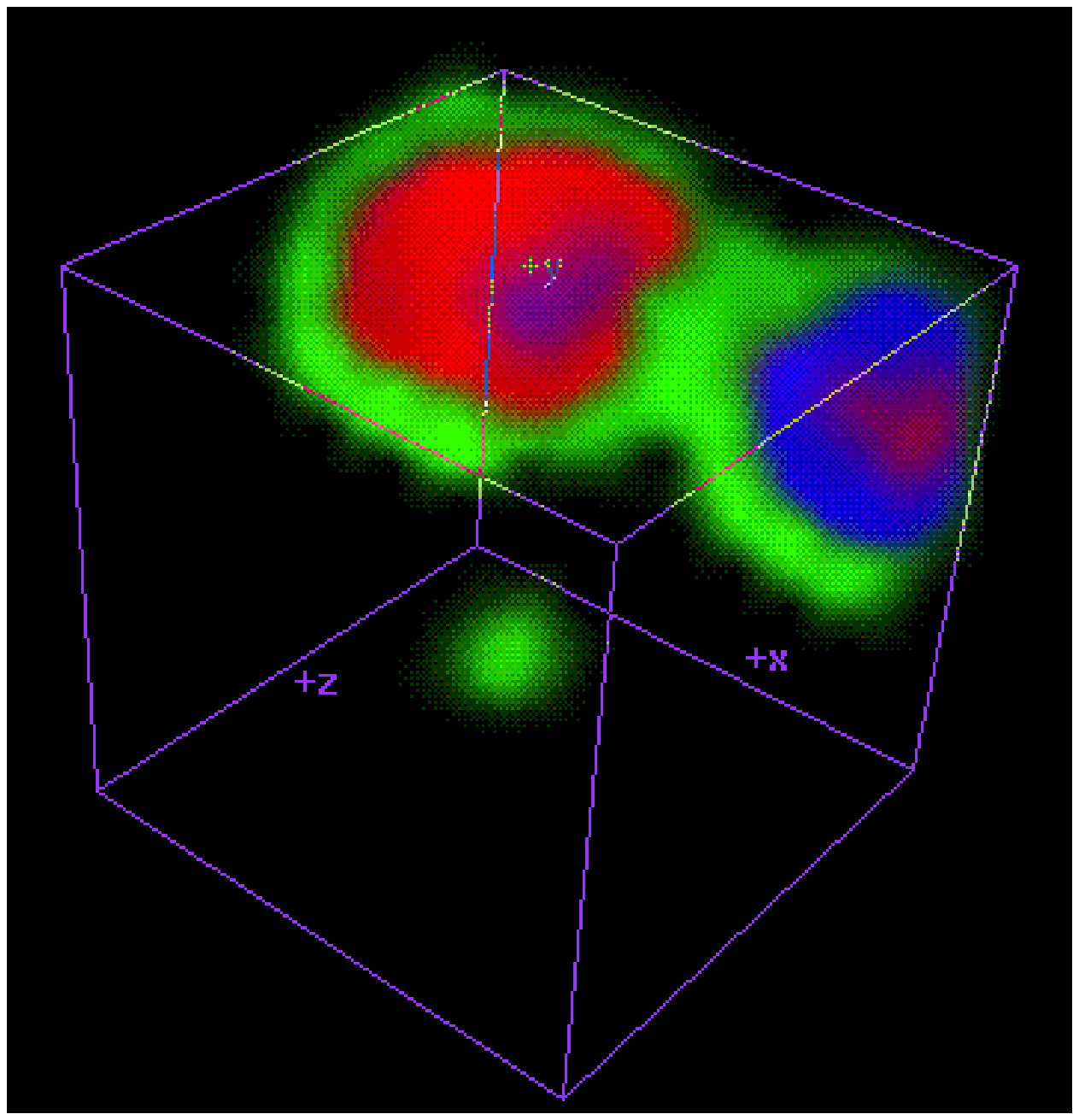,height=6.3cm,width=6.3cm} &
\psfig{figure=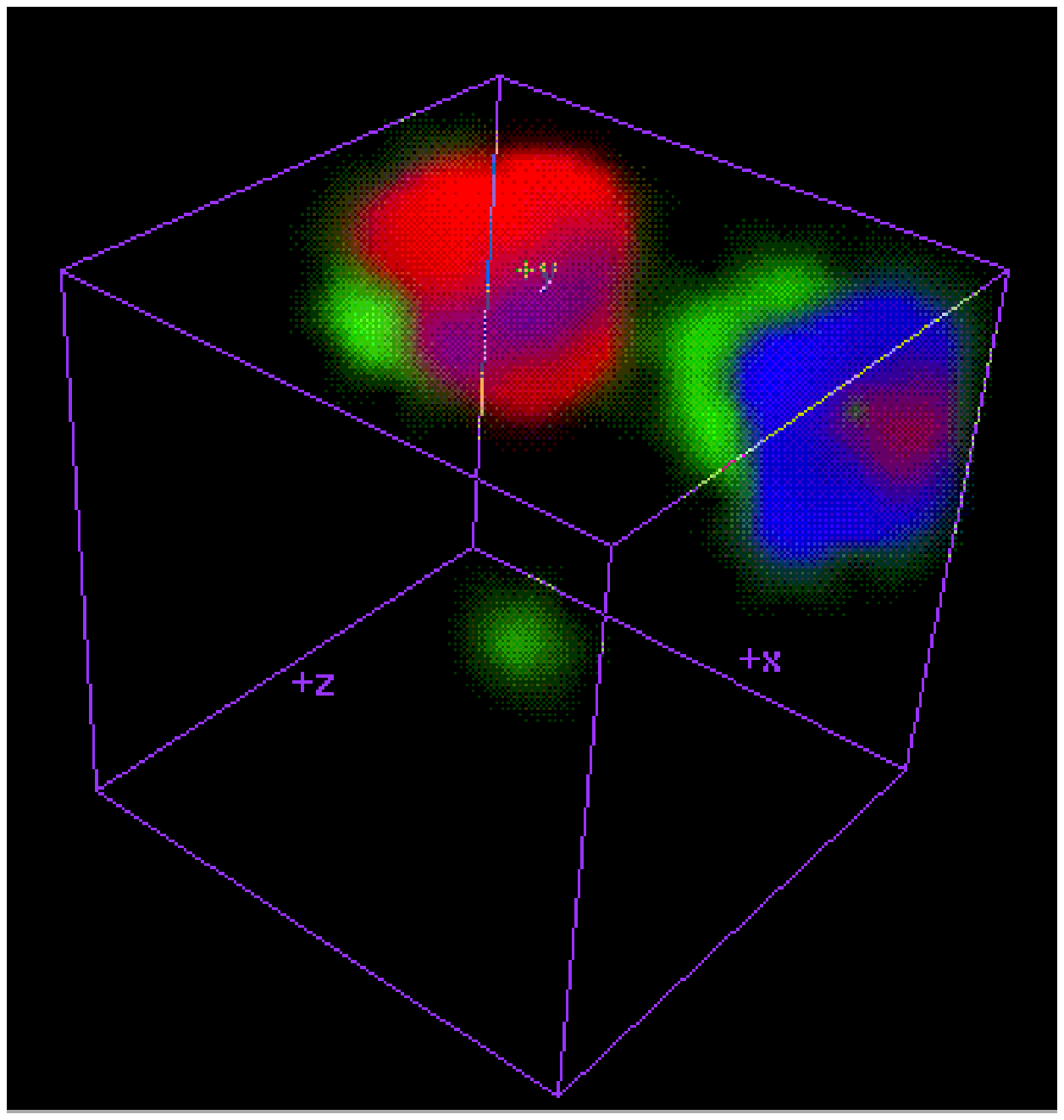,height=6.3cm,width=6.3cm} \\
\end{tabular}
\vspace{-1mm}
\caption{
Cooling history for one time-slice of a gauge field
configuration of $SU(3)$ theory from a non-quenched simulation
in the confinement phase.
The dark and medium grey shades represent regions of positive and negative 
topological charge density respectively; 
the light grey tone surrounding them marks the local quark condensate. 
See the discussion in the text.
}
\label{hist}
\end{figure}

For sufficiently cooled configurations the operator $\bar \psi \psi$
provides an independent measure of the instanton extension and
shape. When the topological density profile is given
by  $q_\rho(x)\sim  \rho^4 \, (x^2+\rho^2)^{-4}$ (as for a classical
instanton) the corresponding quark zero-mode has a profile 
$\bar \psi \psi_\rho (x) \sim \rho^2 \, (x^2+\rho^2)^{-3}$. 
In order to estimate $\rho$ we fitted a convolution of the functional form 
$C_{\bar\psi\psi q}(d) 
\propto \sum_x \bar\psi\psi_\rho(d-x) q_\rho^2(x)$ 
to our data points. 
This was evaluated after 11 cooling steps where the configurations are 
dilute and well-defined instanton sizes can be expected.
Our fit yields $\rho = 1.8$
in lattice spacings. This corresponds to $\rho \simeq .4$ fm
which is consistent with the estimate based
on the $qq$-correlator.
The result is in agreement with other estimates in the literature 
and supports the conclusion
that instantons 
in the presence of dynamical (sea) quarks are somewhat larger than in pure 
Yang-Mills theory.

We now visualize the local quark condensate and the 
topological density on individual gauge field configurations. 
In Fig.~\ref{hist} one time-slice of a typical $SU(3)$ configuration from 
the non-quenched simulation of the confinement phase is shown. This is a
nontrivial $Q=0$ configuration containing one instanton-antiinstanton
pair. The four pictures illustrate how the extension and shape
of the fermionic cloud changes with cooling from the full 
quantum field theory towards a (semi)classical one. We remind that the 
fermionic contribution to the gluonic action 
is neglected in the force driving the gauge field relaxation.
We display the region of positive topological density
with $q(x) > 0.003$ by dark grey shades 
and that with negative density $q(x) < -0.003$ by a medium grey-tone.
The local quark condensate is made visible  by a light grey shade whenever 
a threshold for $\bar \psi\psi (x) > 0.066$ is exceeded. 
By analyzing dozens of lattice field configurations we found the
following picture.
The topological charge is hidden by quantum fluctuations and
becomes visible only by cooling of the gauge fields. For
0 cooling steps no structure can be seen in $q(x)$ or $\bar \psi\psi (x)$ 
on individual configurations. Our results 
discussed above prove that this
does not mean the absence of correlations on the uncooled ensemble. 
After 5 cooling steps clusters of
nonzero topological charge density and local quark condensate are resolved.
Both, the localization of the local condensate around instantons and 
the existence of valence clouds between instantons and antiinstantons,
is visible in this stage of cooling.
During the next 5 cooling steps 
the valence cloud of chiral condensate connecting 
a neighboring instanton-antiinstanton pair becomes gradually weaker.
At approximately 10 cooling steps the local
quark condensate 
forms almost separated clouds coinciding with the
(anti)instantons.
Under further cooling instanton and antiinstanton shrink and finally disappear. 
Due to the Wilson action they  turn   
into dislocations strongly violating the lattice equation of motion.

Beside the quark condensate $\bar \psi \psi$ the quark charge density 
$\psi^{\dagger}\psi$ represents an interesting quantity. 
Similarly  to  QED it reflects the polarization of the QCD 
vacuum into virtual quarks and antiquarks. 
The local structure of this polarization cloud and its correlation 
with topological objects is a challenging issue. 
Without external sources the net vacuum polarization has to be zero. 
We know from the previous discussion of 
$\bar \psi\psi(x)$ that $\psi^{\dagger}\psi(x)$  
can only be clustered where the topological density clusters.
The interesting point is 
the distribution of the sign of $\psi^{\dagger}\psi(x)$. In Fig.~8 
we display a 
non-quenched SU(3) configuration on the same lattice size as above.
Here a tadpole improved action 
for the gluons has been used, both for the
generation of the (pseudofermionic) Monte Carlo configurations
and in the (purely bosonic) cooling. The  couplings in the improved 
bosonic part
of the action were chosen to yield approximately the same lattice 
constant and did not lead to a qualitative change of the
physical results described before. 
After 25 mild cooling sweeps, the chiral condensate 
is still seen to be correlated with the topological charge density. 
The configuration of Fig.~8 
consists of a single instanton sitting at the periodic boundary.
As expected, the quark charge density is concentrated on the topological 
cluster. An interesting feature  exhibited by  the plot, is 
a characteristic change of 
sign of $\psi^{\dagger}\psi$ inside the domain with $q(x)\neq0$. 
This  gives an insight into the local vacuum polarization mechanism of QCD. 
Pairs of virtual quarks and antiquarks are created locally 
predominantly in the  regions of nontrivial topology. 
%
\begin{figure}[!th]
\begin{tabular}{cc}
{\large Chiral Condensate}   & {\large Charge Density} \\

\vspace{-19mm} & \\
\epsfxsize=6.3cm\epsffile{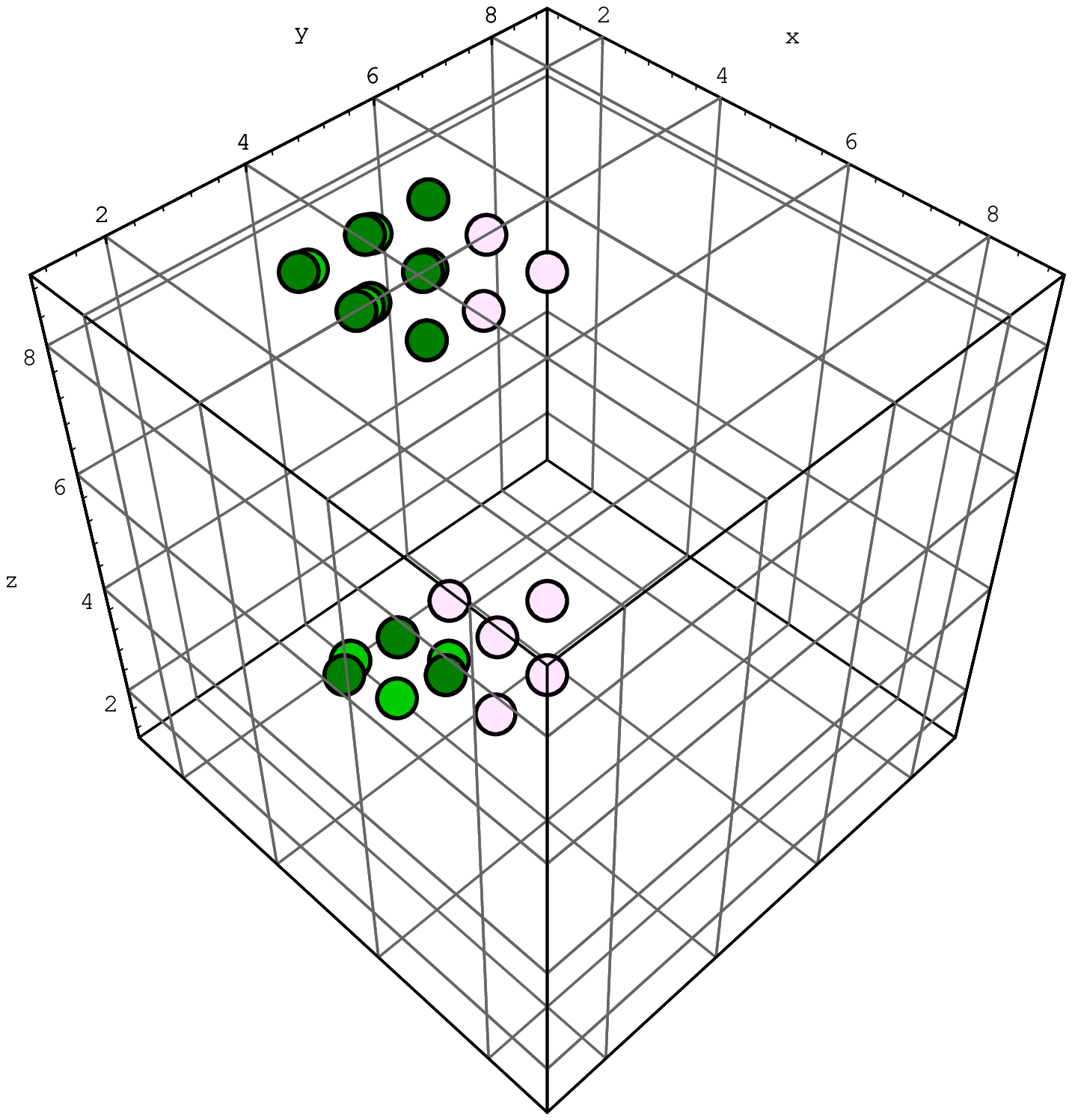}  & 
\epsfxsize=6.3cm\epsffile{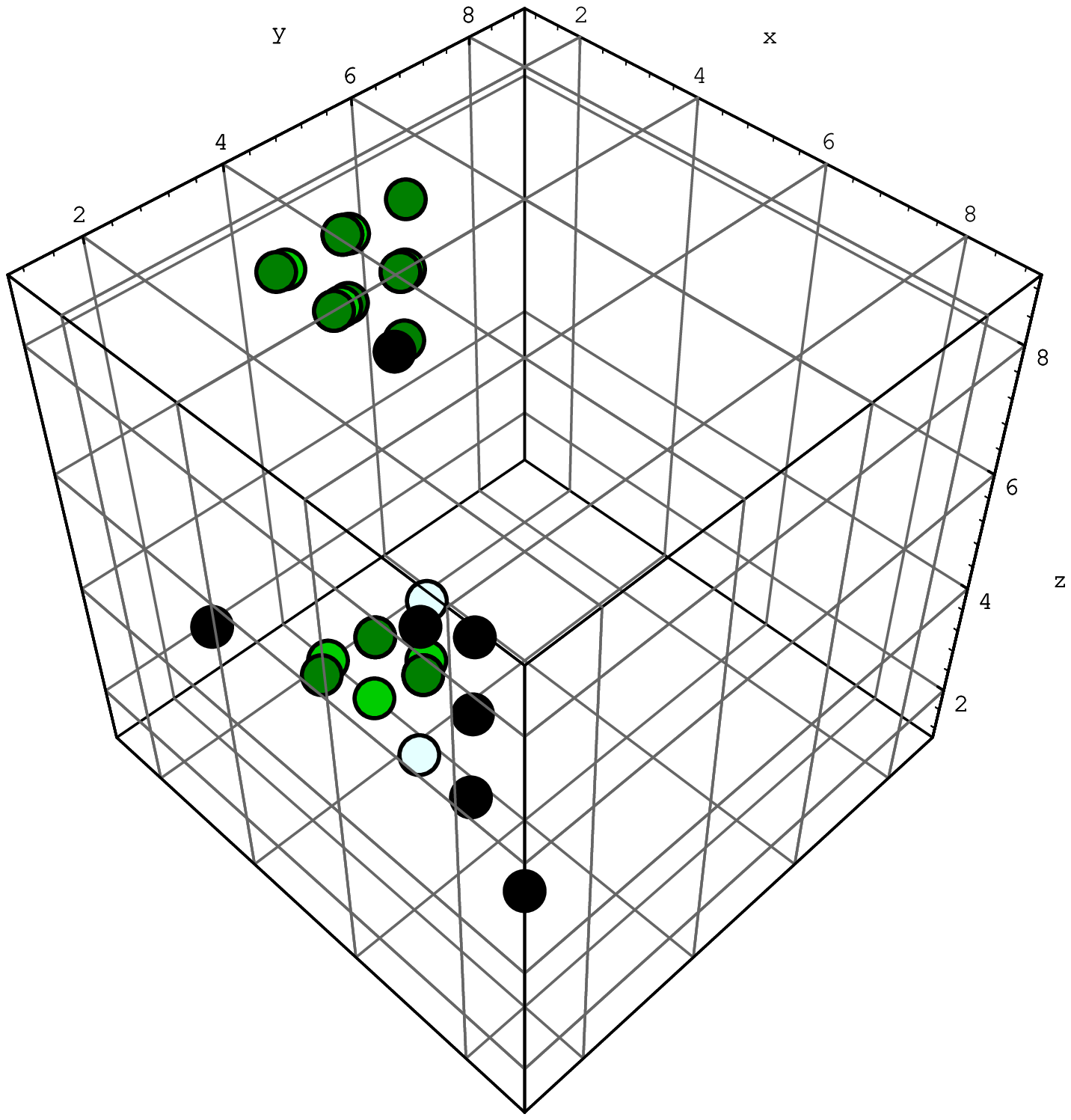 }  \\
\end{tabular}
\vspace{-14mm}
\caption{
Comparison of the local chiral condensate  and the 
quark charge density on a non-quenched $SU(3)$ configuration after 25 
cooling steps. The grey  dots represent the topological charge density. In 
the left plot the light  dots give the local condensate. In the 
right plot the dark  and light dots show the quark charge density 
which fluctuates in sign. 
  }
\label{charge}
\end{figure}

\section{Conclusion}

We have calculated the local values of topological charges and monopole
currents in the maximally Abelian gauge 
and directly displayed them with the help of computer graphics.
After a few cooling sweeps one
observes clearly that instantons are accompanied by monopole loops.
We have evidenced that, by averaging over space-time, a nontrivial
correlation exists already on individual gauge fields. 
This is true also in the deconfinement phase although the structure
of the monopole currents in this case is completely different 
from the structure in the confinement phase. 

Using a renormalization group based
classically perfect action it was shown
that, after elimination of quantum fluctuations of 
$O(a)$ size
by constrained minimization of this action (minimal smoothing), 
clustering of action and topological charge occurs on a scale of
several lattice spacings. These clusters are correlated with  monopole
currents. After smoothing has been applied 
it becomes visible that all strong color
fields  are  locally (approximately) selfdual or antiselfdual.
However, a parametrization of the smoothed configurations in terms of
classical (anti)instantons seems not possible for 
the dense medium of topological excitations
in the confinement phase.  
Since the smoothed configurations contain relatively few 
Abelian monopoles the topological content of the
monopole-instanton relation 
in the Euclidean path integral seems to be accessible for detailed
investigation.
This is necessary in order to identify the role of instantons for
monopole condensation leading to
confinement and to point out the 
role of monopoles for chiral symmetry breaking.

It has turned out  that correlations of monopole currents on one hand with
the topological density and the local
quark condensate on the other 
are rather insensitive with respect to the  
cooling or smoothing technique.
This, taken together with the result of our visualizations of individual 
gauge fields,
leads us to conjecture that the long distance properties of the
topological density distribution 
of gauge field configurations being
relevant for confinement {\it and}
chiral symmetry breaking are encoded in the pattern of
monopole currents. 

We have further demonstrated the local correlation between  
instantons (becoming visible by cooling) on one hand and the local 
quark condensate
and the quark charge density on the other. The correlation functions have 
been evaluated for gauge field
configurations from a non-quenched simulation of the confinement phase,
before cooling as well as with cooling. For our ensemble of configurations,
the correlation function between
topological charge density and local condensate has an extent  
of approximately two lattice spacings which is not very sensitive to 
the amount of cooling. A fit of this correlation with a convolution 
of 
the topological density with $\bar \psi \psi (x)$  leads to an
instanton size of about $0.4$ fm consistent with other methods. 
This instanton radius is larger than for quenched $SU(3)$ gauge theory
due to the effect of sea quarks. 

As expected, the visualization of the local condensate and the quark
charge density shows that both are concentrated to a large extent
on top of the (anti)instantons as soon as the latter become identified
by cooling. More interesting is what we can learn from the cooling history, 
showing quantitatively the change of a quantum field  
to its semiclassical background. When the effect of dynamical quarks is
neglected in  the gluonic cooling, 
the valence cloud of chiral condensate connecting 
a neighboring instanton-antiinstanton pair gradually disappears before,
under further cooling, (anti)instantons shrink and dwindle as dislocations. 
This demonstrates that not just the instantons but also the interpolating 
gauge field is important for the propagation of quarks on the instanton 
liquid background.  These observations give  direct insight into the local
interplay of chiral symmetry breaking and nontrivial topological structure.
Further examination of the fermionic vacuum structure has exhibited that the 
quark charge density fluctuates locally giving thus evidence for 
quark pair creation inside instantons.
It must be emphasized that these fermionic results have been obtained
on a  rather small lattice with finite quark mass. No attempt to
extrapolate towards the thermodynamic and chiral limits has been made.

Most of the statistical correlation functions turned out to be generically 
independent 
of the smoothing technique and of the choice of the action. This indicates
that they are 
a characteristic also for the quantum ensemble of field configurations. 

\section*{ Acknowledgments}
This work was partially supported by FWF under Contract No. P11456-PHY. 
S.~T. would like to thank M.~Feurstein for stimulating discussions.
E.-M.~I., H.~M. and M.~M.-P. gratefully acknowledge the
kind hospitality of T.~Suzuki and all other
organizers of the {\it 1997 Yukawa International Seminar (YKIS'97)}
on {\it Non-Perturbative QCD - Structure of the QCD Vacuum}
at the Yukawa Institute for Theoretical Physics of Kyoto University.
%
%

\end{document}